\documentclass[iop]{emulateapj}
\usepackage{graphicx}
\usepackage{setspace}
\usepackage{natbib}
\usepackage{color}
\usepackage{amsmath,amssymb}
\usepackage{times}
\usepackage{aas_macros}

\begin{document}

\title{Broken and unbroken: the Milky Way and M31 stellar haloes}
\author{A.J Deason\altaffilmark{1,2,3}, V. Belokurov\altaffilmark{3},
  N.W. Evans\altaffilmark{3}, K.V. Johnston\altaffilmark{4}}

\altaffiltext{1}{Department of Astronomy and Astrophysics, University
  of California Santa Cruz, Santa Cruz, CA 95064, USA;
  alis@ucolick.org}
\altaffiltext{2}{Hubble Fellow}
\altaffiltext{3}{Institute of Astronomy, Madingley Rd, Cambridge, CB3
  0HA, UK}
\altaffiltext{4}{Department of Astronomy, Columbia University, New York, NY 10027, USA}

\date{\today}

\begin{abstract}
We use the \cite{bullock05} suite of simulations to study the density
profiles of $L^{*}$-type galaxy stellar haloes. Observations of the
Milky Way and M31 stellar haloes show contrasting results: the Milky
Way has a `broken' profile, where the density falls off more rapidly
beyond $\sim 25$ kpc, while M31 has a smooth profile out to 100
kpc with no obvious break. Simulated stellar haloes, built solely by the accretion of dwarf galaxies, also exhibit this
behavior: some haloes have breaks, while others don't. The presence or
absence of a break in the stellar halo profile can be related to the
accretion history of the galaxy. We find that a break radius is
strongly related to the build up of stars at apocentres. We relate
these findings to observations, and find that the `break' in the Milky
Way density profile is likely associated with a relatively early
($\sim 6-9$ Gyr ago) and massive accretion event. In contrast, the
absence of a break in the M31 stellar halo profile suggests that its
accreted satellites have a wide range of apocentres. Hence, it is
likely that M31 has had a much more prolonged accretion history than
the Milky Way.
\end{abstract}

\section{Introduction}
A diffuse envelope of stars surrounds the Milky Way Galaxy. This halo
of stars only contributes a meager few percent to the total light, but
comprises the oldest, and most metal-poor stars in the Galaxy. The
dynamical timescales in the radial range of the halo stars ($\sim
10-100$ kpc) are long, so these stars can preserve their initial
conditions. It is widely recognized that by studying the phase-space
and chemical properties of the stellar halo we have the opportunity to
unravel the accretion history of our Galaxy.

The density profile of the Milky Way stellar halo has been studied
extensively over the past decade. Early work extending out to $25-30$
kpc found that the halo follows an oblate, single power-law
distribution with minor-to-major axis ratio $q \sim 0.5-0.8$, and
power-law index $\alpha \sim 2-4$ (e.g. \citealt{yanny00};
\citealt{newberg06}; \citealt{juric08}). More recent work probing
further out in the stellar halo has found that the stellar density
falls-off more rapidly beyond $\sim 25-30$ kpc with a power-law index
of $\alpha_{\rm out} \sim 4-5$ beyond the break radius
(e.g. \citealt{watkins09}; \citealt{deason11b};
\citealt{sesar11}). One could argue that the density distribution can
just as easily be described by an Einasto profile
(\citealt{einasto89}), which allows for a steeper fall-off at larger
radii, without resorting to a break. However, whether a broken
power-law or Einasto profile are preferred, in both cases there exists
a characteristic scale (break radius or effective radius) which
deserves a physical justification. Is a break radius a ubiquitous
feature of stellar haloes? Unfortunately, the low surface brightness
of stellar haloes inhibits the study of individual galaxies beyond the
local group. However, our nearest neighbor, M31, provides a useful
comparison.

Early studies called into question whether a stellar halo even exists
in M31: out to $R \sim 30$ kpc the stellar density profile seems to be
a continuation of the M31 bulge (e.g. \citealt{pritchet94};
\citealt{durrell04}). However, more recent work extending to larger
projected distances (out to $R \sim 100-200$ kpc) does indeed find
evidence for a metal-poor halo component with a power-law index of
$\alpha \sim 3.3$ (e.g. \citealt{raja05}; \citealt{irwin05}; \citealt{ibata07}; \citealt{courteau11}; \citealt{gilbert12}). However, there is no evidence for a break in the M31
stellar halo profile: the density profile follows a continuous single
power-law all the way from $R \sim 30$ kpc to at least $R \sim 90$
kpc.

Of course, any measure of the smooth, underlying stellar halo density
profile (if indeed, it does exist), is hampered by the presence of
substructure. A wealth of substructure has now been discovered in both
the Milky Way (e.g. \citealt{ibata95}; \citealt{newberg02};
\citealt{belokurov06}; \citealt{juric08}; \citealt{belokurov07}) and
M31 (e.g \citealt{ibata01}; \citealt{ferguson02}; \citealt{gilbert07};
\citealt{mcconnachie09}). These discoveries have strongly affirmed the
theoretical predictions from numerical simulations that the majority
of the stellar halo is built up from the accretion products of
satellite galaxies.

Numerical simulations of stellar haloes have developed substantially
over the past few years. \cite{bullock05}, hereafter BJ05, presented a
suite of 11 stellar haloes built entirely from the disrupted of
accreted satellites. More recently, \cite{cooper10} used the Aquarius
simulations and a dark matter particle `tagging' method to produce
stellar haloes in a fully cosmological context. \cite{bell08} studied
the `lumpiness' of main sequence turn-off (MSTO) stars in the Milky
Way stellar halo, and found consistency with the \cite{bullock05}
simulations. This led the authors to suggest that the Milky Way
stellar halo is consistent with being built up entirely by
accretion. However, \cite{xue11} and \cite{deason11b} find a somewhat
`smoother' stellar halo when traced by blue horizontal branch (BHB)
stars. These observations perhaps suggest accretion may not be the
only formation mechanism of the stellar halo. Recent hydrodynamic
cosmological simulations postulate that some fraction of the inner
stellar halo is made up of stars formed \textit{in situ}
(e.g. \citealt{zolotov09}; \citealt{font11}). However, stars can
diffuse relatively quickly in configuration space (as opposed to
velocity space), so the \textit{spatial} structure of the stellar halo
can be smooth, even if it is built up from merging and
accretion. Thus, a relatively smooth stellar halo could also signify
an early accretion history.

In this paper we study the density profiles of stellar haloes drawn
from the BJ05 suite of simulations. We investigate the origin of
`broken' density profiles, and in particular, address why some haloes
have an obvious break (e.g. the Milky Way) and why some do not
(e.g. M31). We know that accretion is at least an important (if not
the sole) contributor to the stellar halo. By comparison with the BJ05
simulations we hope to link the observed stellar halo profiles of
local galaxies to their accretion histories.

The paper is arranged as follows. In Section 2 we briefly describe the
BJ05 simulations. In Section 3 we study the stellar halo density
profiles of the 11 haloes drawn from the BJ05 simulations. In Section
4 we investigate the origin of broken halo profiles and we discuss the
implications for the Milky Way and M31 stellar haloes in Section
5. Finally, we summarize our main findings in Section 6.

\section{Bullock and Johnston simulations}

The Bullock \& Johnston simulations (see BJ05 for full details), are a
suite of 11 high resolution stellar haloes built up from the accretion
of dwarf galaxies onto a Milky Way-like potential. The parent galaxy
is represented by a time-dependent, analytical potential consisting of
bulge, disk and halo components. The accretion history of each halo is
randomly drawn within the context of a $\Lambda$CDM Universe; each
accretion event was assigned a binding energy and orbital eccentricity
drawn from the orbital distributions of satellites observed in
cosmological simulations of structure formation. High resolution
N-body simulations were then run to track the evolution of dark matter
particles in each accretion event. At the present day ($z=0$),
individual haloes are built up from the disruption of $N > 100$ dwarf
galaxies.

The baryonic component of each dwarf galaxy is followed by assuming that the cold gas inflow tracks the dark matter accretion rate\footnote{This assumption holds prior to infall into the parent halo; the star formation is truncated soon after each satellite is accreted.}. The gas mass is then used to determine the instantaneous star formation
rate and to track the buildup of stars within each halo. The stellar matter in each dwarf galaxy is associated with the most tightly bound material in the halo. Variable mass-to-light ($M/L$) ratios for each particle are assigned in order to produce realistic dwarf galaxy stellar profiles; the $M/L$ is chosen so
that the luminous matter in the infalling satellites initially follows
a King model embedded within an Navarro-Frenk-White (NFW) dark matter
potential. The King profiles were chosen to specifically reproduce structure (e.g. sigma vs. core relations) observed for dwarf satellites today. The time dependent chemical content of the dwarf galaxies is also included in the models (see \citealt{robertson05} and \citealt{font06} for more details). The stellar population models include enrichment from Type Ia and Type II supernova, as well as stellar winds. A physically motivated supernova feedback prescription is used to reproduce the local dwarf galaxy stellar mass-metallicity relation. The rate of chemical enrichment is calculated analytically, and the abundances of H, He, Fe, and the $\alpha$-elements O and Mg are tracked in the simulations.

BJO5 verified that their models are able to reproduce a number of
observational constraints, such as the number and distribution of
structural parameters of the Milky Way's satellite
population. Furthermore, the resulting stellar haloes have a similar
luminosity ($\sim 10^9 L_\odot$) and density profile to the Milky Way
($\alpha \sim 3.5$), and contain a similar amount of substructure
(\citealt{bell08}). The scatter between the 11 simulated stellar
haloes is due to their different accretion histories; we will explore
the effect of accretion history on the stellar halo density profiles
in the following sections.

\begin{figure*}
  \centering
  \includegraphics[width=7in, height=7in]{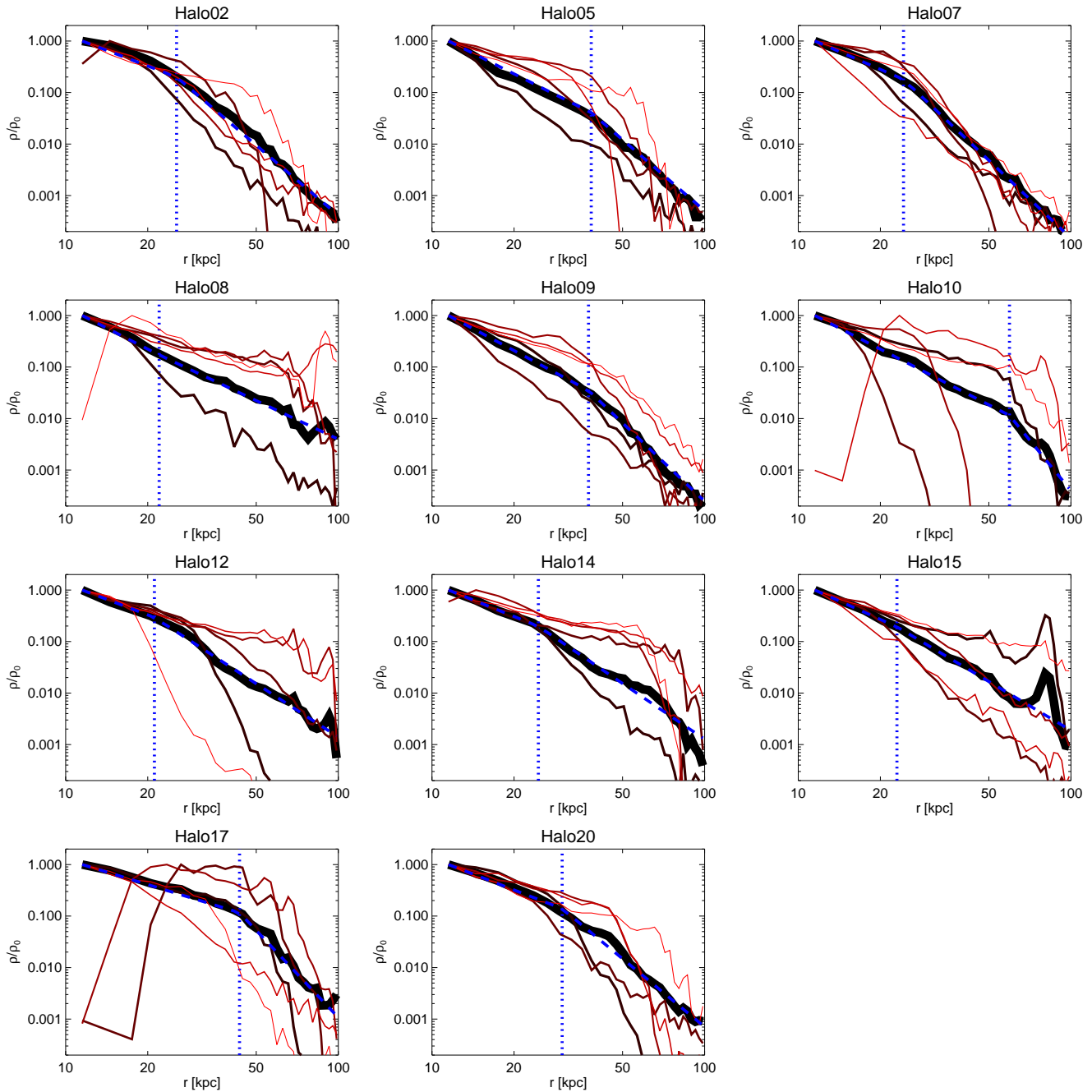}
  \caption[]{\small Radial density profiles of the 11 BJ05 stellar
    haloes. The thick black line shows the overall profile for the 15
    most massive satellites that contribute to the stellar halo within
    100 kpc. The thinner lines show the profiles for the 5 most
    massive accreted subhaloes. The fraction (by mass) contributed to
    the stellar halo decreases from thicker black to thinner red
    lines. The blue dashed line shows the best-fit BPL profile for the
    stellar halo, and the vertical dotted line indicates the
    corresponding break-radius.}
   \label{fig:profs}
\end{figure*}

\begin{table}
\begin{center}
\renewcommand{\tabcolsep}{0.25cm}
\renewcommand{\arraystretch}{1.5}
\begin{tabular}{|  c c  c  c  c  c|}
\hline
ID(BJ05) & SPL-$\alpha$ & BPL-$\alpha_{\rm in}$ & BPL-$\alpha_{\rm out}$ &  BPL-$r_{b}$  & $P_{\rm KS}$ \\
  & & & & [kpc]  & \\
\hline
02(1) & 3.4 & 2.0 & 4.5 & 26 & 0.071\\
05(2) & 3.2 & 2.7 & 4.6 & 38 & 0.020\\
07(3) & 3.6 & 2.3 & 5.0 & 24 & 0.007\\
08(4) & 2.5 & 2.7 & 2.5 & 22 & 1.000\\
09(5) & 3.2 & 2.8 & 5.1 & 38 & 0.001\\
10(6) & 3.0 & 2.7 & 6.5 & 60 & 0.020\\
12(7) & 3.0 & 2.0 & 3.4 & 21 & 0.939\\
14(8) & 3.2 & 2.1 & 3.6 & 25 & 0.985\\
15(9) & 2.9 & 2.4 & 3.1 & 23 & 0.853\\
17(10)& 2.9 & 1.6 & 5.6 & 43 & 0.031\\
20(11)& 3.1 & 2.1 & 4.4 & 30 & 0.207\\
\hline
\end{tabular}
  \caption{\small The global properties of the 11 BJ05 stellar
    haloes. We give the halo ID from galaxia (see http://galaxia.sourceforge.net/) and BJ05 in parenthesis, the best-fit single power-law slope,
    the best-fit broken power-law slopes and break radii and the KS
    test (SPL versus BPL) probability.}
\label{tab:props}
\end{center}
\end{table}

\section{Stellar halo density profile}

We investigate the density profile of halo stars belonging to the
eleven parent haloes discussed in BJ05. Table 1 in BJ05 shows that the
majority (80-90 \%) of the stellar halo is built up from the 15 most
massive accreted satellites. For this reason, we only consider the
contribution to the stellar halo from the 15 largest satellites. We
fit single power-law (SPL) and broken power-law (BPL) density profiles
to stellar haloes between 10-100 kpc. To fit the density profiles, we
compute the density in radial bins and use the \textsc{mpfit}
\textsc{idL} routine (\citealt{mpfit}) to find the minimum Chi-square
SPL or BPL density profile. The best-fit parameters for the density
profiles are given in Table \ref{tab:props}. In addition, we also fit
profiles for stars belonging to \textit{individual} accretion
events. Provided the accretion event is not too recent (less than 4.5
Gyr ago, cf. Figure 3 in \citealt{johnston08}), a BPL is a good
description of the density profile (see e.g. Fig. \ref{fig:vr_apo}).

The density profile for each stellar halo is shown by the thick black
lines in Fig. \ref{fig:profs}. The dashed blue line shows the best-fit
BPL profile for the stellar halo, and the break radius for this fit is
shown by the vertical dotted line. The thinner lines show the profiles
for the 5 most massive accreted satellites. The fraction (by mass)
contributed to the stellar halo (between 10-100 kpc) decreases from
thicker black lines to thinner red lines.

Fig. \ref{fig:profs} shows that a broken power-law profile is
generally a good description of the stellar halo density profiles,
where the stellar density falls off more rapidly beyond the break
radius. Such `broken' stellar haloes are also seen in the stellar
haloes of \cite{cooper10} (see also BJ05). However, it is worth noting
that some stellar haloes can just as easily be described by a single
power-law.

\section{Origin of broken profiles}
\label{sec:origin}

The stellar halo in the BJ05 models is a superposition of stars
stripped from several satellite galaxies. Thus, to understand the
origin of the global stellar halo profile, we must first investigate
the profiles of stars belonging to individual accretion events. The
density profile of stars stripped from an individual satellite galaxy
is well-described by a BPL (see e.g. Fig. \ref{fig:vr_apo}). What
causes this BPL profile?

For each star particle, we find the apocentre and pericentre of its
orbit. Assuming a spherically symmetric, stationary potential, we can
estimate the apocentre and pericentre from the following equation (see
\citealt{binney87}, Chapter 3):
\begin{equation}
\label{eq:perap}
u^2  +  \frac{2[\Phi(1/u)- E]}{L^2} = 0. 
\end{equation}
Here, $u=1/r$ and the roots of this equation give the apocentre and
pericentre. This equation can be solved for each star particle given
the potential and particle properties defined at redshift $z=0$. This
is a good approximation to the orbital properties of the stars at the
\textit{time of stripping}. However, the orbital properties may have
evolved since the time of accretion onto the parent halo. In Appendix
\ref{sec:app}, we verify some of our deductions made from the halo
star properties at redshift $z=0$ by tracing back the orbital
histories of the accreted satellites.
 
\begin{figure}
  \centering
  \includegraphics[width=3in, height=4in]{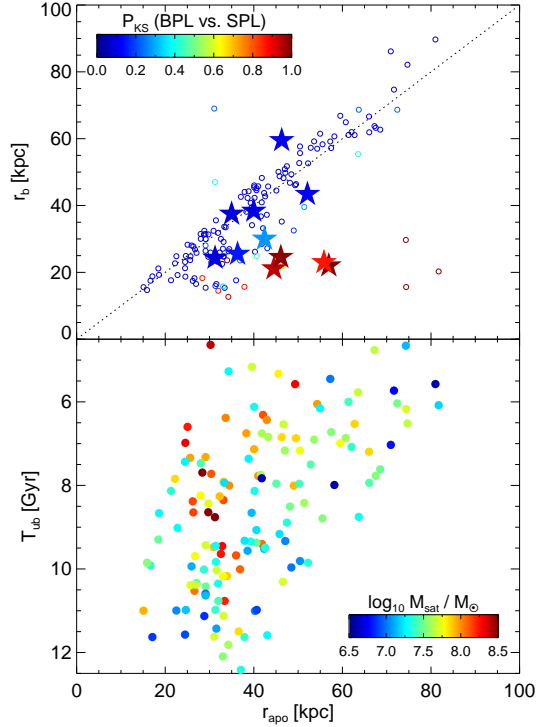}
  \caption[]{\small \textit{Top panel:} The break radius of the
    stellar density profile against the average apocentre of the star
    particle orbits. The circle symbols indicate individual accreted
    satellite remnants from all 11 stellar haloes, and the star
    symbols indicate their global values. The colors indicate the
    probability that the best-fit single power-law density profile is
    drawn from the same distribution as the best-fit broken power-law
    density profile (i.e. using a two-sided KS test). The red colors
    indicate profiles which do not have a strong break in the stellar
    halo density. The dotted line shows a one-to-one
    relation. \textit{Bottom panel:} The average time of stripping for
    each accreted satellite against the apocentre of the
    orbits. Satellites accreted at early times have smaller apocentres
    as the physical size of the parent halo is smaller. The colors
    indicate the satellite (stellar) mass. More massive satellites are
    more affected by dynamical friction, so can sink to the center of
    their parent halo more quickly than less massive satellites.}
   \label{fig:rb}
\end{figure}

\begin{figure*}
  \centering
  \includegraphics[width=5.76in, height=3.2in]{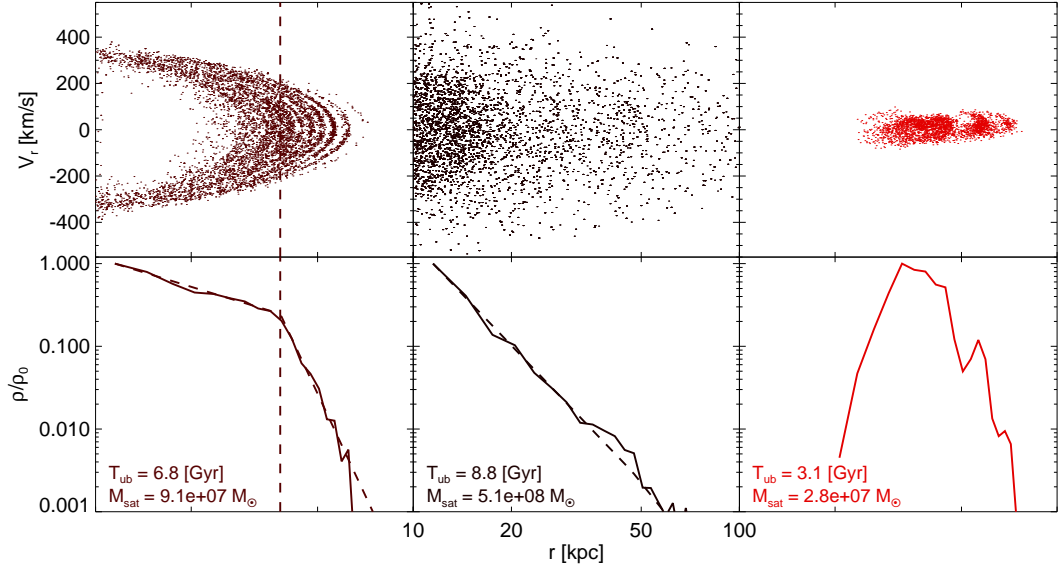}
  \caption[]{\small Examples of individual satellite contributions to
    the global stellar halo. Three satellites accreted/disrupted at
    different times in Halo05 are shown. The top panels show their
    radial velocities as a function of radius and the bottom panels
    show their density profiles. The dashed-line is the best-fit
    broken power-law model. In the left-hand panel, there is an
    obvious `break' in the stellar density which coincides with the
    average apocentre of the star particle orbits. Satellites
    disrupted a long time ago ($T_{\rm ub} \gtrsim 9$ Gyr) are often well
    phase-mixed and follow a single power-law profile. Recent
    accretion events ($T_{\rm ub} \lesssim 4.5$ Gyr) are often unrelaxed and
    have poorly defined (i.e. non-smooth) density profiles.}
   \label{fig:vr_apo}
\end{figure*}

From the inferred orbital properties of the stars, we can find the
average apocentre and pericentre of the stars belonging to an
individual satellite accretion event. As the star particles can have
different masses (see BJ05 for details), we compute the mass weighted
average. In the top panel of Fig. \ref{fig:rb}, we show the average
apocentre of star particles belonging to one satellite against the
best-fit break radius for the density profile of these stars. The
satellites from all eleven parent haloes are shown in this plot. We
also show the average properties for the global stellar haloes with
the star symbols. There is a clear correlation between apocentre and
break radius; here, the dotted line indicates a one-to-one relation.

The colors in the top panel of Fig. \ref{fig:rb} indicate the
\textit{strength} of the break. To characterize the break strength we
perform a two-sided KS test on the best-fit SPL and BPL density
models. Weak breaks will have a high probability of being drawn from
the same distribution as the SPL profile (i.e. $P_{\rm KS} \sim 1$),
while strong breaks will have a low probability (i.e. $P_{\rm KS} \sim
0$). One can see that the points deviating the most from a one-to-one
relation between apocentre and break radius tend to have the weakest
breaks.

The bottom panel of Fig. \ref{fig:rb} shows the average time of
(stellar) stripping for each satellite ($T_{\rm ub}$, in Gyr) against
the average apocentre at $z=0$. Satellites accreted at early times
have smaller apocentres as the physical size (and mass) of the parent
galaxy is smaller. Therefore, breaks give an indication of
\textit{when} a satellite was accreted and/or disrupted. However, a
further complication is dynamical friction, as more massive satellites
can sink to the center of the parent halo relatively quickly. The
colors in the bottom panel of Fig. \ref{fig:rb} indicate the satellite
(stellar) mass. More massive satellites tend to have smaller
apocentres (at stripping) even if they are accreted relatively
recently.

In Fig. \ref{fig:vr_apo}, we show the radial velocity structure of
three accreted satellites in Halo05. The stripped material from the
satellite in the left-hand panel shows an obvious break in the density
profile which coincides with the apocentres of its star particle
orbits. The middle panel shows a (massive) satellite accreted a long
time ago ($\sim 9$ Gyr ago) which is now well mixed in phase-space
and can be described by a single power-law density profile
(cf. \citealt{johnston08}). Finally, the right-hand panel illustrates
a relatively recent accretion remnant for which the density profile is
poorly defined. In general, satellites accreted between 4.5-9 Gyr ago
follow a BPL density profile. Satellites accreted a long time ago
$T_{\rm ub} \gtrsim 9$ Gyr, can be well mixed in phase-space, and recent
accretion events ($T_{\rm ub} \lesssim 4.5$ Gyr ago) are unrelaxed and their
density profiles are poorly defined.

\begin{figure*}
  \centering
  \includegraphics[width=4.8in, height=3.84in]{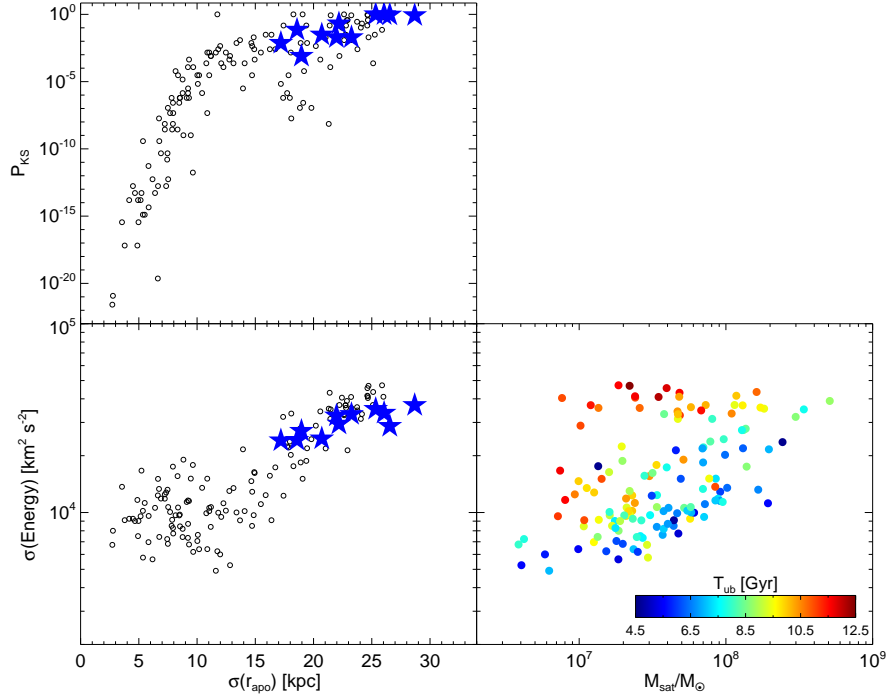}
  \caption[]{\small \textit{Top panel:} The `strength' of the broken
    power-law density model against the dispersion of star particle
    apocentres. The strength of the break is characterized by a KS
    test between single and broken power-law density models. A high
    probability that the best-fit single power-law model is drawn from
    the same distribution as the best-fit broken power-law model
    indicates a weak, if not non-existent, break in the stellar
    density profile. The black circles are individual satellites and
    the blue stars indicate the global stellar halo
    properties. \textit{Bottom-left panel:} The dispersion in energy
    of star particles against the dispersion in star particle
    apocentres. A large range of star particle energies leads to a
    large dispersion in their apocentres, and hence a weaker break in
    the stellar density profile. \textit{Bottom-right panel:} The
    dispersion in energy of star particles against satellite
    mass. Higher mass satellites have a larger range of particle
    energies. The colors indicate the time of stripping.}
   \label{fig:strength}
\end{figure*}

We have found that the satellite apocentre sets the scale of the break
radius, but what sets the `strength' of the break? In the top panel of
Fig. \ref{fig:strength} we show the break strength (defined by a
two-sided KS test between SPL and BPL profiles) against the spread of
apocentres for stars belonging to an individual satellite. We find
that stronger breaks have a smaller spread in particle apocentres. The
spread in apocentres is related to the energy distribution of the star
particles; the bottom left-hand panel of Fig. \ref{fig:strength} shows
that the dispersion in energy of an accreted remnant is strongly
correlated with the spread in particle apocentres. Note that these
relations also hold for the global stellar halo properties (shown by
the blue star symbols). Here, the spread in apocentre is calculated
from all 15 satellite accretion remnants.

The bottom right-hand panel of Fig. \ref{fig:strength} shows that more
massive satellites have a larger spread in energy (as expected analytically, see e.g. \citealt{johnston98}). The colors indicate
the average time of stripping for each satellite. Satellites accreted
a long time ago ($ T_{\rm ub} \gtrsim 9$ Gyr) can have a large spread in
energy, regardless of the satellite mass (cf. phase-mixed example in
middle-panel of Fig. \ref{fig:vr_apo}).

\section{Implications for global stellar halo properties}
\begin{figure*}
  \centering
  \includegraphics[width=5.76in, height=3.84in]{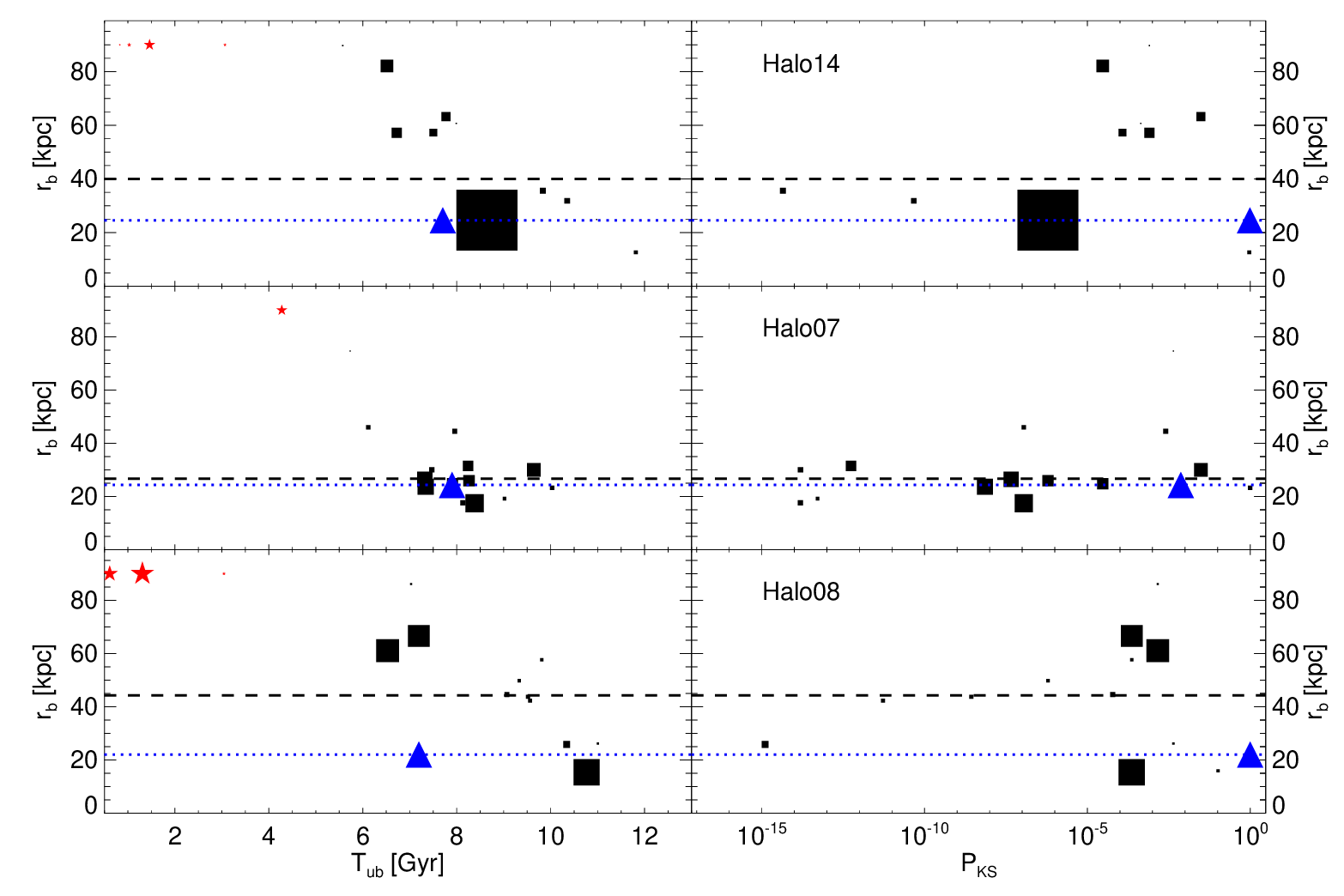}
  \caption[]{\small Examples of three stellar haloes. The black
    squares show the break radii of the 15 most massive accretion
    events against disruption time (left panels) and strength of break
    (right panels). The sizes of the square symbols indicate the
    fractional contribution of each satellite to the overall stellar
    halo (larger squares are more massive satellites). The blue dotted
    line and blue triangles indicate the overall stellar halo
    properties. The red stars indicate recent accretion events ($
    T_{\rm ub} \lesssim 4.5$ Gyr) where the density profiles of the remnants
    are poorly defined. These are placed at $r_b = 90$ kpc for
    illustrative purposes. The dashed black line indicates the
    \textit{mass-weighted} break radius. This is calculated by
    computing the weighted mean of all 15 satellites' break
    radii. Haloes 14 and 07 (top and middle panels) can both be
    described by a broken power-law stellar density profile (albeit
    Halo14 has a relatively weak break: $P_{\rm KS} \sim 0.99$ and
    $\alpha_2-\alpha_1=1.5$) . In the latter case the accreted
    satellites all have similar apocentres (and hence break-radii)
    whereas the former halo is dominated by one massive
    satellite. Halo08 (bottom panel) has no obvious break ($P_{\rm KS}
    \sim 1$ and $\alpha_2-\alpha_1=-0.2$) and its accreted satellites
    have a wide range of apocentres.}
   \label{fig:global}
\end{figure*}

\begin{figure}
  \centering
  \includegraphics[width=3.33in, height=5in]{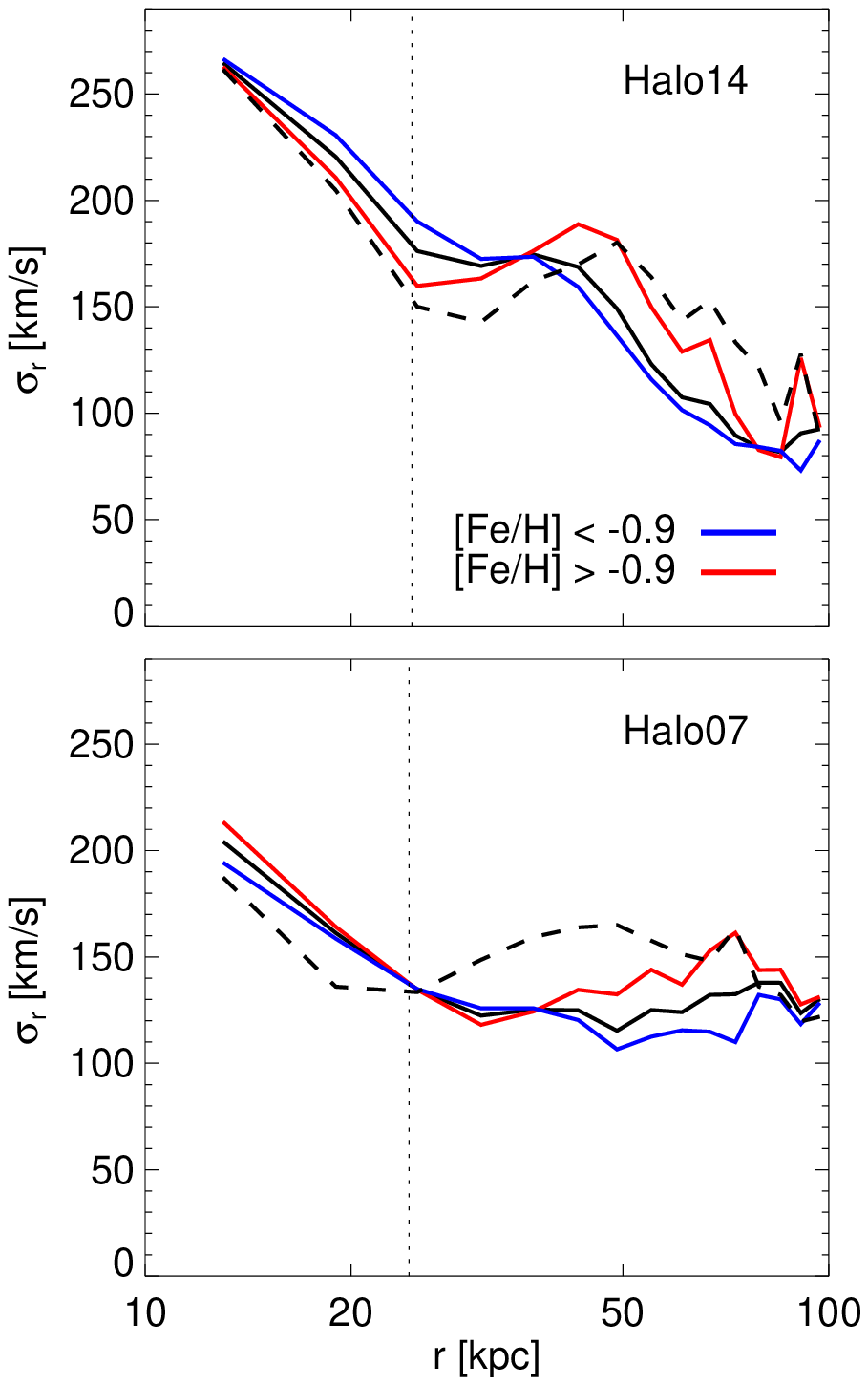}
  \caption[]{\small Radial velocity dispersion profile for two example
    haloes. The black lines show the overall profiles, the red/blue
    lines show the profiles for star particles with metallicity
    greater than/less than $[\mathrm{Fe/H}] = -0.9$. The dashed lines
    show the profiles for the stars belonging to the most massive
    satellite accreted by the galaxy. The dotted lines indicate the
    break radii of each halo. The stellar halo of Halo14 is dominated
    by one massive satellite whilst Halo07 is built up from several
    similar mass satellites with similar apocentres. The
    metal-rich(er) velocity dispersion profile for Halo14 shows a more
    pronounced 'dip' near the break radius. There is little difference
    between the metal-rich(er) and metal-poor(er) profiles for Halo07,
    indicating that the break is not dominated by a single, massive
    accretion event.}
 \label{fig:sigr_met}
\end{figure}

In the previous section, we found that the density profile of stars
stripped from an individual satellite is strongly related to the
orbital structure of the stars. The average apocentre of the star
particle orbits corresponds to the break radius of the best-fit BPL
density profile. Furthermore, the strength of the break depends on the
spread in energy (and hence apocentres) of the stars. In general, more
massive satellites have a larger energy spread. However, very early
accretion events ($ T_{\rm ub} \gtrsim 9$ Gyr ago) which are well mixed in
phase-space can also have a large spread in energy, regardless of the
initial satellite mass.

We now relate these findings to the global stellar halo
properties. The global stellar halo is a superposition of individual
satellite contributions. Therefore, the overall stellar density break
radius reflects the individual satellite contributions. For example, a
strong break ensues when the accreted satellites all have similar
apocentres (e.g. Halo07, middle panel of Fig. \ref{fig:global}). The
halo can also be dominated by one massive satellite (e.g. Halo14, top
panel of Fig. \ref{fig:global}), and the overall break radius reflects
the apocentre of this massive satellite. In Fig. \ref{fig:global} we
show the best-fitting break radius for individual satellite
contributions versus average time of stripping (left-hand column) and
strength of break (right-hand column). The size of the black squares
illustrates the satellite mass (larger symbols for more massive
satellites). The blue triangles indicate the global parent stellar
halo properties and the blue dotted line indicates the parent stellar
halo break radius. We also show the `mass-weighted' global break
radius by the black-dashed line. This is computed from the individual
satellite break radii weighted by their mass. The top two rows
illustrate the two different scenarios mentioned above. In Halo14 the
overall stellar halo density profile is dominated by one massive
satellite, whilst in Halo07 several satellites (of comparable) mass
have very similar break radii.

Are we able to discern between these two scenarios?
Fig. \ref{fig:sigr_met} shows the radial velocity dispersion profiles
for these two haloes. The black line is the overall profile, and the
red and blue colors illustrate the profiles for metal-rich(er) and
metal-poor(er) stars (arbitrarily split at the average metallicity of
the stellar haloes $[\langle \mathrm{Fe/H} \rangle]= -0.9$). The
profile for the most massive satellite contributing to the stellar
halo (between 10-100 kpc) is shown by the black dashed line. In both
Halo07 and Halo14 there are `dips' in the velocity dispersion profile
near the break radius (indicated by the vertical dotted line). This is
a result of the low radial velocities of the star particles at the
apocentres of their orbits. However, in Halo14 this dip is much more
pronounced in the higher-metallicity stars. This is because a massive
satellite, which is therefore relatively metal-rich, dominates the
break in Halo14. One can see that the profile for this massive
satellite (dashed black line) closely follows the profile for the most
metal-rich stars in the halo. There is little difference between the
metal-rich and metal-poor material in Halo07. This is because all of
the satellites in Halo07 (low and high mass) have very similar
apocentres, and there is no massive, dominating accretion remnant.

Finally, we consider the case where there is no obvious break in the
stellar density profile. In this case, the accretion events may have a
wide range of apocentres (e.g. Halo08, bottom row of
Fig. \ref{fig:global}). As the apocentre is related to the time of
accretion, this suggests that the accretion history has been
prolonged. The bottom row of Fig. \ref{fig:global} shows that there
have been significant satellite disruption events over a wide time
interval: from $\sim 11$ Gyr ago to recent events only $\sim 1$ Gyr
ago. In comparison, Halo14 and Halo07 show little activity in the last
6 Gyr.

In summary, the stellar halo density profile, and in particular the
presence or absence of a break in the stellar density, can give us an
important insight into the accretion history of a galaxy.

\subsection{The Milky Way and M31}

In the last few years, several groups have found that the Milky Way
stellar halo has a `broken' density profile (e.g. \citealt{bell08};
\citealt{sesar11}; \citealt{deason11b}), with $r_b=20-30$ kpc and
$\alpha_{\rm in}=2.3-2.8$ and $\alpha_{\rm out}=3.8-4.6$. Furthermore,
\cite{deason11a} found a dip in the line-of-sight\footnote{In the
  radial range of the \cite{deason11a} study, 10-50 kpc, the
  line-of-sight velocity is a good approximation for the radial
  velocity} velocity dispersion profile of the \textit{metal-rich(er)}
($[\mathrm{Fe/H}] > -2$) blue horizontal branch stars (see Figure 6 in
\citealt{deason11a}). This dip in velocity dispersion occurs between
20-30 kpc, in the approximate radial range where there is a break in
the stellar density profile. 

Our conclusions in the previous sections suggest that these
observations may have several important implications:
\begin{itemize}
\item The relatively close-by break radius of the Milky Way
  ($r_b=20-30$ kpc) is likely caused by accreted satellites with
  small apocentres (at the time of stripping).
\item The cold velocity dispersion profile near the break radius
  provides further evidence for the break-apocentre connection.
\item A metal-rich bias in the velocity dispersion profile suggests
  that the break in the Milky Way stellar halo could be dominated by a
  (relatively) \textit{massive}
  satellite(s).
\item The density profile has only been measured out to 50 kpc, so we cannot
  rule out more recent accretion events with larger
  apocentres. However, the relatively smooth stellar density profile
  out to 50 kpc (\citealt{deason11b}) suggests \textit{early} accretion events and then somewhat quiescent evolution.
\end{itemize}

Recently, \cite{gilbert12} measured the density profile of the
M31 stellar halo out to 100 kpc (see also earlier work by
\citealt{raja05}, \citealt{irwin05}, \citealt{ibata07} and \citealt{courteau11}). These authors find that the
stellar distribution can be described by a SPL with $\alpha \sim 3.3$,
and there is no evidence for a break in the density profile. As we
found in the previous section, no obvious break in the global stellar
halo profile indicates that the stars have a wide range of apocentres,
which suggests that M31 has had a much more prolonged accretion
history than the Milky Way (see e.g. Halo08). This deduction, that M31
appears to have experienced a much more active accretion history than
the Milky Way, is in good agreement with other independent
observations of the two galaxies: Relative to the Milky Way, M31 has a
more disturbed disc (e.g. \citealt{brown06}), a larger bulge
(e.g. \citealt{pritchet94}; \citealt{durrell04}), a younger and more
metal-rich halo population (e.g. \citealt{irwin05};
\citealt{kalirai06}), and more numerous tidal streams and surviving
satellite galaxies (e.g. \citealt{koch08}; \citealt{richardson11}).

\section{Conclusions}

The stellar haloes of the Milky Way and M31 have conflicting density
profiles: the Milky Way has a broken profile, whereby the stellar
density falls off more rapidly beyond a break radius ($r_b \sim 25$
kpc), whereas, the stellar halo of M31 shows no obvious break and can
be described by a single power-law. In light of these recent
observations, we have studied the density profiles of stellar haloes -- built solely by the accretion of dwarf galaxies -- drawn from the \cite{bullock05} suite of simulations, with the aim of understanding the contrasting stellar haloes of our local galaxies.

We summarize our conclusions as follows:

\bigskip
\noindent
(1) The simulated haloes often have `broken' stellar halo profiles,
where the density falls off more quickly beyond the break radius.
However, some haloes do not have an obvious break, and their density
distribution can be described by a single power-law (e.g the Milky Way
versus M31). 

\bigskip
\noindent
(2) In the BJ05 simulations, the global stellar halo is a
superposition of accretion products from the $\sim 15$ massive
accretion events. The density profiles of the stars that once belonged
to an individual satellite follow a broken profile, where the break
radius corresponds to the average apocentre of the stars. However,
material belonging to recently stripped satellites ($T_{\rm ub} \lesssim 4.5$
Gyr) have ill-defined density profiles, and very ancient accretion
events ($T_{\rm ub} \gtrsim 9$ Gyr) are often well-mixed in phase-space.

\bigskip
\noindent
(3) The location of the break in the stellar density is linked to the
time of accretion/stripping and the mass of the satellite. Satellites accreted
at early times have smaller apocentres as the physical size of the
parent halo is smaller. More massive satellites can have smaller
apocentres (and hence break radii) at the time of stripping, as the
satellite can spiral into the center of the galaxy via dynamical friction.

\bigskip
\noindent
(4) The strength of the break in the stellar density depends on the
spread of apocentres of the stars. For the global stellar halo, the
break strength depends on the range of apocentres of its accreted
components. Individual satellite accretion remnants can also have
varying break strengths. More massive satellites have a larger spread
in energy (and hence apocentres). Also, stars belonging to satellites
accreted a long time ago can have a wide spread in energy as they are
well mixed in phase-space today.

\bigskip
\noindent
(5) The global stellar halo density profile depends on the accretion
history of the galaxy. An obvious break in the overall density profile
suggests that: (a) the accreted satellites have similar apocentres; or
(b) one massive satellite dominates the stellar density in the
applicable radial range. These two scenarios could be distinguished
observationally. The radial velocities of stars near apocentre are
very low, thus the radial velocity dispersion is also low
(i.e. shell-type structures). In case (b) above, the apocentres of
stars that once belonged to a massive, and hence metal-rich, satellite
dominate the break. Thus, a `dip' in the velocity dispersion profile,
the signature of a build-up of apocentres, will be more pronounced in
the \textit{metal-rich(er)} material. Conversely, if the accreted
satellites all have similar masses (and similar apocentres), then we
would expect no metallicity bias.

\bigskip
\noindent
(6) The profiles of some haloes show no obvious break in the stellar
density. Often, these haloes have a prolonged accretion history
whereby satellites are accreted over a wide range of timescales ($\sim
0-10$ Gyr ago). Thus, the average apocentres of the accreted
satellites, whose material now makes up the stellar halo, also have a
wide range of values.

\bigskip
\noindent
Our investigation into the simulated BJ05 stellar haloes has provided
some important insights into the accretion histories of the Milky Way
and M31. The absence of a break in the M31 stellar halo suggests that
this galaxy has had a prolonged accretion history, where the accreted
satellites had a wide range of apocentres. However, the strong break
in the Milky Way stellar halo (at $r_b = 20-30$ kpc), in addition to
the presence of a shell-type feature in the relatively metal-rich BHB
stars, suggests that the stellar halo break is dominated by the
apocentre of a (relatively) massive satellite.

\section*{Acknowledgments}
We thank an anonymous referee for valuable comments. AJD is currently supported by NASA through Hubble Fellowship grant HST-HF-51302.01, awarded by the Space Telescope Science Institute, which is operated by the Association of Universities for Research in Astronomy, Inc., for NASA, under contract NAS5-26555.

\label{lastpage}
\bibliography{mybib}

\begin{appendix}
\section{Relating satellite properties at accretion to $z=0$ halo stars}
\label{sec:app}
In this section, we verify some of our deductions from the star
particle properties at $z=0$, by tracing back the accreted satellites whose stripped stellar material make up the
stellar haloes today. 

\begin{figure*}
  \centering
  \begin{minipage}{0.49\linewidth}
    \centering
    \includegraphics[width=8.7cm, height=8.7cm]{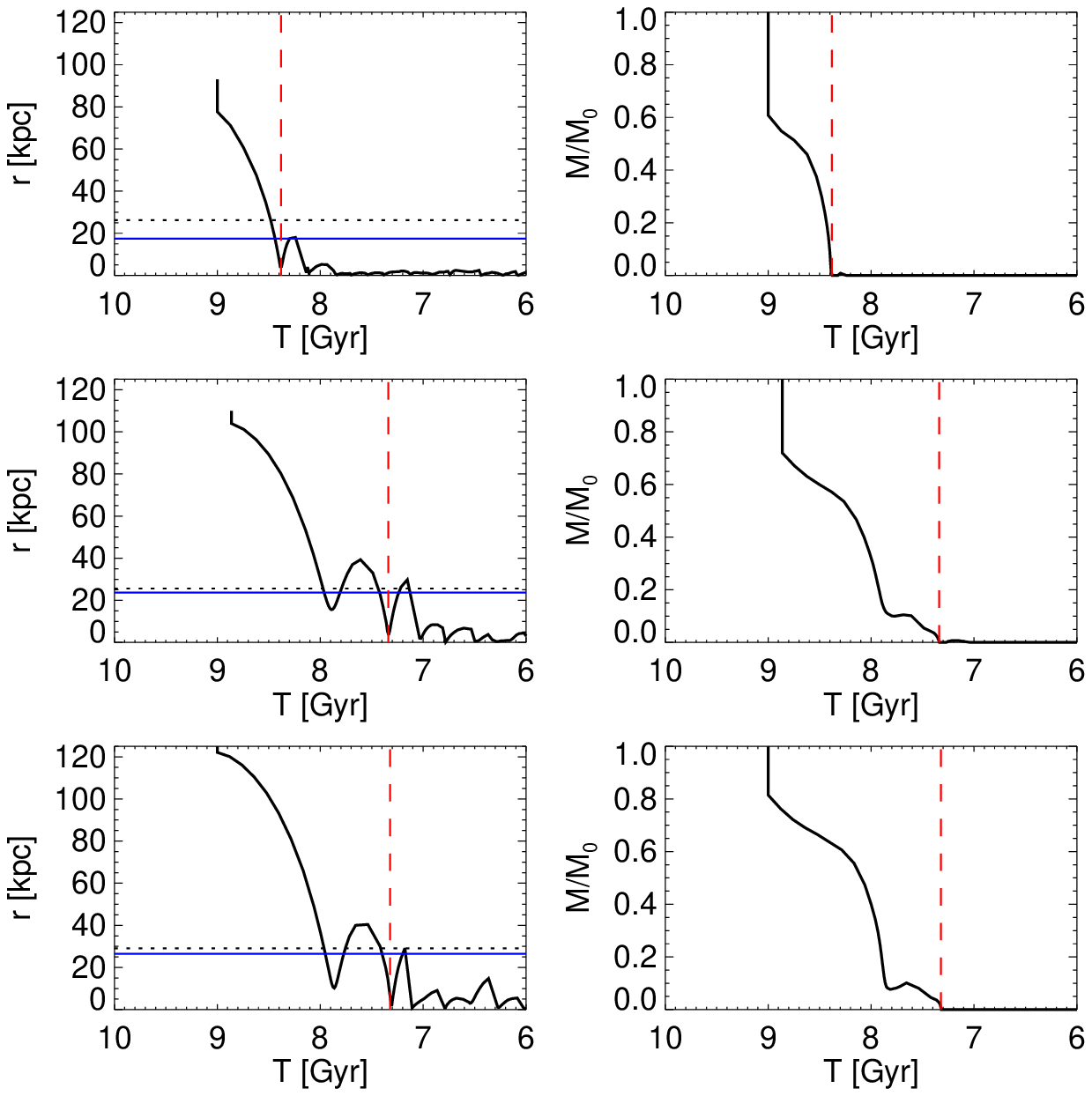}
  \end{minipage}\hfill
  \begin{minipage}{0.49\linewidth}
    \centering
    \includegraphics[width=8.7cm, height=8.7cm]{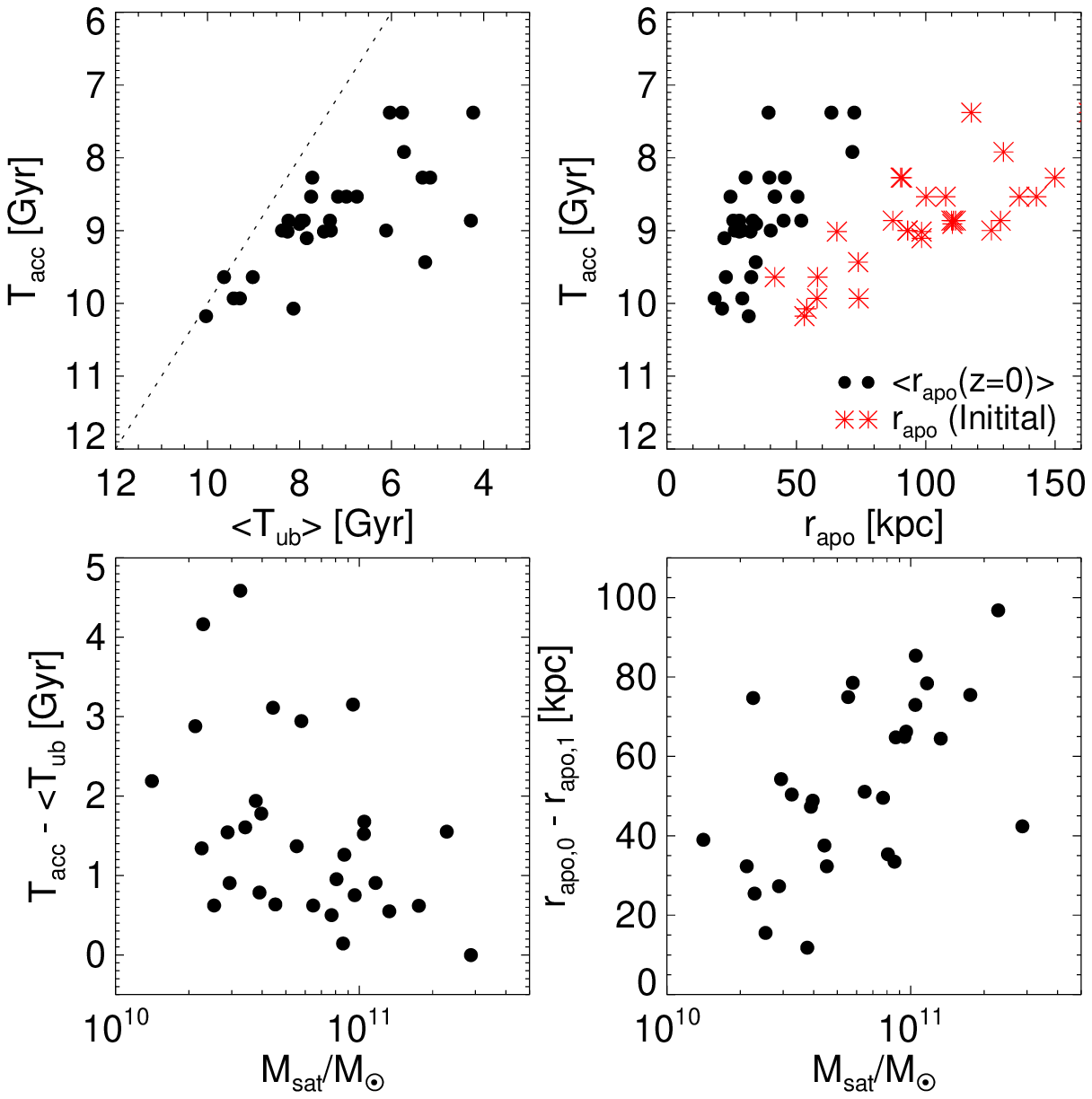}
  \end{minipage}
  \caption{\small \textit{Left hand panels:} Three example satellite
  orbits from Halo07. The radius (left columns) and mass (right
  columns) are shown as a function of time. The satellites are
  followed from their initial apocentre (at accretion) until the
  satellite is completely disrupted. The dashed red line
  shows the average time at which the stars belonging to the satellite
  are stripped. The dotted black lines show the average apocentre of
  the star particles (at stripping) calculated in Section
  \ref{sec:origin}. The solid, blue line shows the estimated break
  radius of the stripped stellar material. Here, we confirm our
  deductions in the main text; the break radius is coincident
  with the apocentre of the satellite near stripping. \textit{Right
  hand panels:} The top-left panel shows that the average time at
  which stars are stripped from the satellite is correlated with the
  time at which the satellite is accreted. The top-right panel shows
  that the accretion time of a satellite is correlated with both the
  initial apocentre (red asterisks), and the final apocentre (filled
  black circles). The bottom two panels show the effect of dynamical
  friction. We show the orbital time (left panel) and reduction in
  apocentre (right panel) against satellite mass. More massive satellites are
  stripped quicker, and have a larger difference between initial and
  final apocentre.}
  \label{fig:trace}
 \end{figure*}

We follow the orbits of satellites from Halo02 and Halo07. In the
left-hand panels of Fig. \ref{fig:trace} we show three examples of
satellites from Halo07. These are the three most massive contributors
to the stellar halo within 100 kpc of this galaxy. In the left columns
we show the evolution of radius with time for the satellites, and the
right columns show the evolution of total satellite mass with time. The
red dashed line indicates the average time when stars become unbound
($<T_{\rm ub}>$) from the satellite. The dotted black line indicates the
average apocentre of the star particles computed in Section
\ref{sec:origin}, and the solid blue line
indicates the approximate break radius in the density profile of the
stripped stellar material. This figure shows that the break radius we
measure at $z=0$ from the stars is indeed coincident with the
apocentre of the host satellite galaxy near the time of stripping. In
Section \ref{sec:origin}, we deduced this by estimating the apocentre
of star particle orbits using Equation \ref{eq:perap}.

In the right-hand panels of Fig.\ref{fig:trace} we relate some of the
initial satellite properties to those we infer from their stripped
stars at $z=0$. The top-left panel of this figure shows the relation
between accretion time (of the satellite) and the average time at
which the star particles become unbound. The stars are
stripped a few Gyr after satellite accretion, but there remains a
positive correlation between these two timescales. The bottom-left
panel shows that the difference between these timescales is related to
the mass of the satellite galaxy. Due to dynamical friction effects,
more massive satellites spend a shorter
amount of time orbiting the galaxy before the stellar material is
stripped. In the top-right panel we show satellite accretion time against
initial satellite apocentre (red asterisks) and average star particle
apocentres at $z=0$ (filled black circles). In Section \ref{sec:origin}, we
noted that larger apocentres correspond to more recent accretion
events as the physical size, and mass of the parent galaxy is larger,
and hence more distant satellites can be captured. The strong trend
between initial apocentre and accretion time reinforces this
statement. Although the trend is slightly weaker, we see that the
apocentres of the stripped stars at $z=0$ are still an indication of
when their host satellite was accreted. The final apocentre of the
satellite (and hence stripped stars) is generally reduced from the
initial apocentre via dynamical friction effects. In the bottom-right
panel of the figure we show that the difference in apocentre radius is
strongly related to the satellite mass.

In summary, we find that our deductions made from the $z=0$ halo stars in the main section of the text,
are verified when we trace the accreted satellite properties back in time.

\end{appendix}

\end{document}